# *Quantum Approximate Optimization Algorithm: Performance on Simulators and Quantum Hardware*

Abyan Khabir Irfan
*Department of Computer Science*
*Cleveland State University*
Cleveland, OH, USA
a.irfan@vikes.csuohio.edu

Chansu Yu
*Department of Electrical and Computer Engineering*
*Cleveland State University*
Cleveland, OH, USA
c.yu91@csuohio.edu

***Abstract*—Running quantum circuits on quantum computers does not always generate "clean" results, unlike on a simulator, as noise plays a significant role in any quantum device. To explore this, we experimented with the Quantum Approximate Optimization Algorithm (QAOA) on quantum simulators and real quantum hardware. QAOA is a hybrid classical-quantum algorithm and requires hundreds or thousands of independent executions of the quantum circuit for optimization, which typically goes beyond the publicly available resources for quantum computing. We were granted access to the IBM Quantum System One at the Cleveland Clinic, the first on-premises IBM system in the U.S. (127-qubit IBM Eagle R3, >50 us coherence time, <0.5% single-qubit error, <2.5% two-qubit gate error, <5% measurement error). This paper explores different optimization methods, techniques, and error mitigation methods to observe how they react to quantum noise differently, which is helpful for other researchers to understand the complexities of running QAOA on real quantum hardware and the challenges faced in dealing with noise.**

## I. INTRODUCTION

Quantum computing offers powerful new ways to solve problems using principles like superposition and entanglement. Unlike classical computers, quantum systems can process many possibilities at once, allowing certain problems to be solved much faster, including optimization problems [1]. However, programming quantum computers is very different from classical systems and comes with challenges, including coherence time and qubit error.

While large-scale quantum computers are still years away, current Noisy Intermediate-Scale Quantum (NISQ) devices can still tackle certain problems by using hybrid quantum-classical methods like variational algorithms, which optimize quantum circuits using classical feedback [2]. Among them, Quantum Approximation Optimization Algorithm (QAOA) and Variational Quantum Eigensolver (VQE) show promise. Studies suggest that the shallow circuit depth of QAOA and VQE helps make them noise-tolerant, as demonstrated through simulations with different quantum noise channels [3].

This paper uses QAOA to compare its performance on a simulator and quantum hardware, and demonstrates the impact of noise/error as well as the potential design choices to avoid them. We used the optimization problem called the Max-Cut as an example, which is a well-known NP-complete problem and one of the key benchmarks for assessing quantum optimization algorithms [4].

We utilized the *ibm_cleveland* system in this study, which has the IBM Eagle R3, a 127-qubit processor. It integrates classical hardware for system control, custom electronics for signal generation, and a dilution refrigerator for low-temperature operation. The quantum processor uses fixed-frequency transmon qubits based on superconducting Josephson junctions, coupled with niobium-based resonators and buses [15]. Designed for high fidelity, the system achieves a coherence time greater than 50 microseconds, single-qubit gate errors below 0.5%, two-qubit gate errors under 2.5%, and measurement errors under 5%. These specifications make the IBM Eagle R3 a suitable platform for running complex quantum algorithms like QAOA, despite the challenges of noise and error rates in real quantum hardware [5].

Our contribution in this paper is to investigate the challenges of running QAOA on real quantum hardware and present our findings in terms of solution quality (e.g., cost or MaxCut value), as well as the evolution of cost and parameter values throughout the optimization process. We experimented with (i) different minimizing methods such as COBYLA, Powell, and CG, (ii) varying depth of the QAOA circuit (denoted as p in this paper), and (iii) pulse-based error mitigation methods.

We made the following observations: First, we found that the CG minimizing method works well on noisy hardware. It is interesting to observe that different minimization methods produce quite different patterns in the cost trajectory throughout the optimization process on both the simulator and the quantum hardware. Interesting fluctuations were also observed in the parameter progression. Second, unlike our expectations, we observed that a higher QAOA circuit depth (p) does not improve performance under the experiment environment and configurations we tested. Third, we found positive results with QAOA on quantum hardware, with error mitigation methods improving our results.

The rest of the paper is organized as follows: Section II introduces the Maxcut problem and explains the QAOA circuit, process, and parameters. Section III goes through our experimental design. Section IV presents results on both the quantum simulator and real quantum hardware. Section V concludes the paper and outlines directions for future work.

## II. BACKGROUND

### A. The Maxcut Problem

This paper uses QAOA to solve the Maxcut problem [4], which is a classical optimization problem where the goal involves dividing a graph's nodes into two groups so that the number of edges that connect these groups is maximized [3]. In Fig. 1 (a), the maxcut is achieved with {0,1,2} and {3,4}, and the maxcut value is evidently 6. When a bit indicates the group each node belongs to, the bitstring 00011 or 11100 achieves the maximum cut, 6. It is an NP-complete problem and can be solved using QAOA. Our implementation is based on the code from Ruslan Shaydulin's GitHub [6].

### B. QAOA

Quantum Approximate Optimization Algorithm (QAOA) is a hybrid quantum-classical algorithm and is used to solve combinatorial optimization problems such as the Maxcut [4]. QAOA alternates between applying two types of quantum operations, the Cost Hamiltonian and the Mixer Hamiltonian. The Cost Hamiltonian encodes the problem to be solved, guiding the quantum system toward better solutions, while the Mixer Hamiltonian introduces transitions that help explore the solution space widely and avoid getting trapped in local minima.

**Cost Hamiltonian**: The Cost Hamiltonian is responsible for encoding the problem's objective function into the quantum state. This was done by applying controlled rotational gates between pairs of qubits corresponding to the edges of the graph, as in Fig. 1(b). In other words, a series of controlled-NOT (CNOT) gates and a rotation around the Z-axis is applied, parameterized by γ, to encode the energy (cost) contribution of each edge. With a large (small)γ, the quantum state is biased strongly (weakly) toward configurations where edges are cut.

Note that a quantum variational algorithm such as QAOA uses a variational quantum circuit that represents a collection of parameterized states to explore, which is called an ansatz. Unlike heuristic and hardware-efficient ansatz, problem-specific ansatz uses problem-specific knowledge in constructing the QAOA circuit to restrict the search space to a specific type for a speedy search [7].

**Mixer Hamiltonian**: The Mixer Hamiltonian explores different possible solutions by applying transformations that shift the quantum state across the solution space. The Mixer Hamiltonian is usually constructed from X-rotations (Pauli-X gates) applied to each qubit as shown in Fig. 1(c). These rotations change the state of the qubits, mixing the possible solutions and allowing the algorithm to search a broader space for the optimal solution.

While the Cost Hamiltonian encodes the problem by marking good solutions with phases, the Mixer Hamiltonian converts phase information into measurable probabilities by rotating the state around the X-axis, changing the latitude on the Bloch sphere. QAOA alternates the two Hamiltonians to solve combinatorial problems [8].

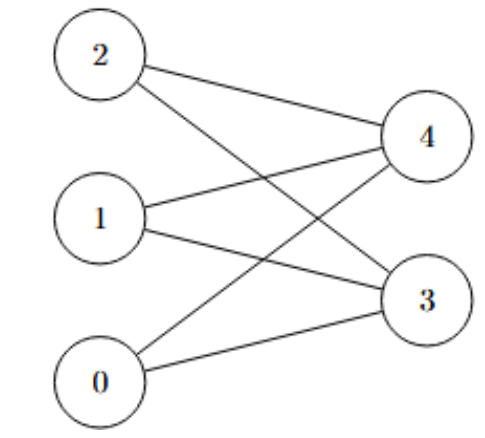

(a) The Maxcut problem graph with five nodes

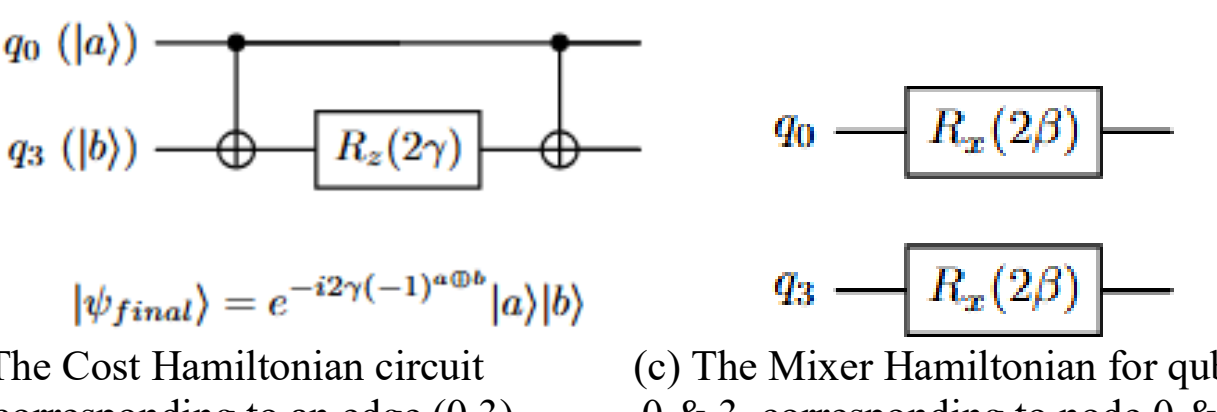

(b) The Cost Hamiltonian circuit corresponding to an edge (0,3)

(c) The Mixer Hamiltonian for qubit 0 & 3, corresponding to node 0 & 3

Fig. 1. The Maxcut problem and its implementation in QAOA.
(In (b), the first CNOT gate calculates the XOR of the two qubits. It is 0 if they belong to the same group and 1 otherwise. The Rz gate creates a phase shift of angle γ (same group) or -γ (different group). I.e., the qubit for a heavily connected node will be more influential in determining the cut pattern, leading to an optimal solution. The second CNOT gate returns the qubits to their original states except for the phase shift.)

### C. The QAOA Process

QAOA begins with an initial state, usually a uniform superposition of all possible solutions, created by applying Hadamard gates to all qubits. The algorithm then alternates the Cost Hamiltonian and the Mixer Hamiltonian. This is repeated for $p$ times for the system to adapt its configuration. This is called the "adiabatic process", where gradually changing conditions allow the system to start in an eigenstate of the initial Hamiltonian but to end in that of the final Hamiltonian [9].

More specifically, for each step $i \in [1, p]$, the quantum state evolves by applying the Cost Hamiltonian, resulting in a phase evolution proportional to a parameter $\gamma_i$ (for those qubits sharing an edge as in Fig. 1(b)), followed by the Mixer Hamiltonian, which induces another phase evolution proportional to a parameter $\beta_i$ (for all five qubits as in Fig. 1(c)). The parameters γ and β are variational parameters that the algorithm optimizes to minimize the expected value of the energy state, thus finding the maximum cut. After the final step, the quantum state is measured, and the result is estimated to be a possible solution to the optimization problem.

Note that the above-mentioned QAOA process is repeated a hundred times to gather probabilities, and the best solution is selected from the measurement outcomes. Using Qiskit's Sampler primitive with the simulator (Aer Simulator) and the real quantum hardware (*ibm_cleveland*), we aimed to observe the impact of "noise"—a common challenge in quantum

computation—on the results obtained from the quantum computer.

### D. Parameter Optimization

**Objective Function**: A classical optimization technique was used to optimize the QAOA parameters (β's and γ's). The objective was to minimize the expectation value of the Cost Hamiltonian, which corresponds to finding the optimal solution to the Maxcut problem.

The Maxcut Energy is computed by taking the average energy (cost) for the Maxcut problem based on measurement outcomes from a quantum circuit. For each bitstring result sampled, it calculates the Maxcut value and averages these values based on their occurrence. The average energy is returned, with a lower value indicating a better solution. The function computes the energy of a given quantum state and returns it as the value to be minimized [6].

**Optimization Methods***:* The parameters were optimized using the minimize function from the *scipy.optimize* library. Various minimizing methods were explored, including COBYLA, Powell, and CG [4].

- Powell is a gradient-free optimization method that does not require the function to be differentiable but does need it to be real-valued. It operates by performing sequential one-dimensional minimization along a set of initial search vectors. In each iteration, Powell calculates a series of conjugate directions and iteratively adjusts the search vectors until the convergence criteria are met.
- COBYLA, on the other hand, is a gradient-free simplex method used for constrained optimization. It approximates the problem iteratively by solving linear programming problems, updating the simplex with each step based on the evaluated values of the objective function. It adjusts a trust region radius for the next iteration, improving the optimization process.
- Conjugate Gradient (CG) can be used in gradient-free and gradient-based forms. The gradient-free form of the CG method estimates search directions using function evaluations rather than exact gradients. By building a sequence of conjugate directions, it can efficiently navigate the parameter space and converge faster. It works well for smooth objective functions but is more noise-sensitive than Powell's method.

These methods are advantageous in noisy environments as they do not require gradient evaluations, relying instead on evaluating the objective function at multiple points determined by conjugate vectors (in Powell) or polynomial functions (in COBYLA) [4].

### E. Error Mitigation Methods

Nonetheless, one of the challenges in quantum computing is to address the noise problem when running on real quantum systems and to come up with noise mitigation methods. Some of them are outlined below, along with how to activate them in IBM's Qiskit Runtime.

**Pre-calibrated two-qubit gates:** IBM compilers can leverage pre-calibrated 2-qubit pulses to implement arbitrary rotations. This can be practically implemented with transpiler parameters of "optimization_level = 3", with "scheduling_method" = 'alap' or 'asap'. Automatically replace sequences of fixed-angle gates with the fastest native pulses.

**Pauli twirling and basis changes:** Twirling, also known as randomized compiling, is a widely used technique for converting arbitrary noise channels into noise channels with a more specific structure. Pauli twirling is a special kind of twirling that uses Pauli operations. It has the effect of transforming any quantum channel into a Pauli channel. This can mitigate coherent noise because coherent noise accumulates quadratically with the number of operations, whereas Pauli noise accumulates linearly. IBM's Runtime provides gate and measurement twirling options that suppress coherent errors, "*estimator.options.twirling.enable_gates = True*". We can enable Pauli (randomized compiling) twirling on two-qubit gates, which symmetrizes errors into stochastic Pauli channels.

**Dynamical decoupling:** Idle qubits accumulate phase noise and crosstalk. Dynamical decoupling inserts sequences of pulses that don't change the overall purpose of the circuit, during idle periods, to refocus errors. For example, the "XpXm" sequence (an X pulse, then its inverse) cancels out low-frequency noise. In Qiskit Runtime, one can enable Dynamic Decoupling on an Estimator or Sampler, which will insert the chosen sequence on any qubit that remains idle for a gap:

```
est = Estimator(mode=backend)
est.options.dynamical_decoupling.enable = True
est.options.dynamical_decoupling.sequence_type = "XpXm"
```

**Pulse Programming:** Pulse programming is the lowest-level form of quantum control available on most superconducting qubit hardware. Instead of working with abstract gates (H, CX, etc.), it works directly with microwave pulses that drive qubit state transitions and timing schedules to control the exact order and overlap of these pulses. In superconducting qubits (like IBM or OQC), gates are compiled into these pulses before execution. Pulse programming allows us to design and optimize them directly.

Pulse-level programming improves quantum computing by implementing multi-qubit interactions directly at the pulse level; we can bypass multiple decomposed gates. Dynamical decoupling sequences can be inserted between operations to reduce decoherence. Crosstalk mitigation can be implemented by shaping pulses and adjusting timing. Custom pulses can exploit specific native hardware pulses. Algorithms like QAOA and VQE can see significant performance boosts when ansatz gates are implemented directly at the pulse level.

Note that the Qiskit Pulse library was available in Qiskit v1.4 but has been deprecated; instead, its functionality is being absorbed in the Qiskit Dynamics as of this writing.

**Fire Opal:** Q-CTRL's Fire Opal quantum solver [17] is designed to extract maximum performance from noisy intermediate-scale quantum (NISQ) devices [16]. It solves, for example, nontrivial (up to 127 qubits) Max-Cut and spin-glass optimization problems at scales previously thought infeasible on gate-model quantum devices, outperforming both quantum annealers (e.g., D-Wave) and other gate-model approaches. Its notable features are:

- AI is used to improve pulse sequences for increased fidelity automatically.
- Fire Opal's error suppression pipeline: The QAOA solver makes use of the execute function's best-in-class error suppression, which raises the standard of individual circuit execution.
- Specialized compilation: To create a shorter circuit and shorter duration, multi-qubit gate operations are parallelized to modify the input circuit. For example, it instructs the transpiler to use built-in fractional RZZ gates and cross-resonance pulses provided by the QPU.
- Pulse-efficient gates: Fire Opal finds recurrent complex gates and optimizes their direct implementation at the pulse level in addition to optimizing the native gate set's implementation. When compared to conventional decomposition, this method reduces the time by half.

## III. Experimental Design

### A. Quantum Backend & Tools

For the simulation and execution of the QAOA circuit, two different quantum backends were used:

- *AerSimulator*: A quantum simulator from Qiskit, used for running the circuit and obtaining results in a classical environment. Note that the *AerSimulator* can be configured to incorporate a *live noise model*, where the noise model of an IBM System. Alternatively, a *Fake Provider* (e.g., FakeManilaV2) can be used. It is a simulated representation of real IBM quantum hardware based on historical calibration data.
- *IBM Quantum Device:* For this project, we used the *ibm_cleveland* quantum backend at the Cleveland Clinic, thanks to the CSU-Clinic agreement. We also used *ibm_sherbrooke* and *ibm_brisbane.* Note that the *IBM Quantum Device* can employ *Fire Opal* to mitigate errors and improve performance.

Note that the live noise model specifies an IBM quantum hardware name in the parameters of the function so that we can import the live noise model to imitate the real-time physical behavior of qubits. Also note that the Fake Providers are used to test circuit transpilation, error rates, and hardware-aware compilation without live QPU access.

### B. QAOA Circuit

The QAOA algorithm was implemented using the *Qiskit* library [11], which provides tools for constructing and running quantum circuits. The QAOA circuit was constructed in the following steps.

**Optimization with Different Minimizing Methods:** We experimented with different minimizing methods in the QAOA algorithm, such as COBYLA, Powell, and CG, observing the spread of noise among the bitstrings on both the simulator and quantum hardware.

**Optimization with different circuit depth, p*:*** We changed the value of p (1 ~ 5) and observed its effect on the complexity of the circuits and the quality of the result.

## IV. Results

Section IV.A discusses the optimization process, followed by experiment results from the simulator (AerSimulator) and IBM Quantum device in sections IV.B and IV.C, respectively. Section IV.B includes simulation results with a live noise model and a Fake Provider to see how much they mimic a real runtime environment. Section IV.D shows IBM backend results with error mitigation methods to see how much they address the noise problem. Section IV.E provides an analysis and explains the limitations of this work.

### A. Optimization process

The optimizer adjusts the parameters θ= (β1, γ1, β2, γ2…, β5, γ5) to minimize the objective function. The optimized parameters were then used to obtain the final QAOA circuit. The next step was to transpile and run the circuit to obtain the counts (bitstrings). For each of the bitstrings in the counts, we obtained the Maxcut value. They are summed to get an average as discussed in section II.D. The progression of the energies over several iterations throughout the optimization process was stored in *numpy* arrays and is plotted. The graph gives insight into how the QAOA parameters evolve during each iteration of the optimization process.

### B. Simulator Results

**COBYLA and Powell:** When using the *AerSimulator*, we used the COBYLA and Powell optimization algorithms. With p = 5, we used the following parameters as initial points (1st half β, 2nd half γ): [2.083, 2.048, 1.792, 1.564, 1.387, 2.281, 5.962, 1.789, 3.563, 5.646]. With COBYLA, the parameters are optimized in 81 iterations. The energy progression is shown in Fig. 2(a). Using the optimized parameters, we constructed and ran the final QAOA circuit. The two most prominent bitstrings (solutions) were 00011 and 11100, as expected, shown in Fig. 3(d).

Optimizing with the Powell algorithm took 1243 iterations, significantly more than COBYLA. As shown in Fig. 2(b), Powell also reached a lower minimum energy. Note that the x-axis range differs between the two plots due to Powell's higher iteration count, and the y-axis range is different as well, with COBYLA's energy varying from -5.5 to -2.5, while Powell's energy ranges from -6 to -1. The energy progression was more haphazard in comparison to COBYLA. With the optimized parameters, the bitstrings obtained from the final QAOA circuit were similar to the ones when COBYLA was used as an optimization method. Except, the bitstring 11100 was more frequently explored.

**COBYLA with circuit depth p**: By decreasing the value of p, we decrease the number of parameters for β and γ, making the circuit less complex and easier to compute, but risking obtaining the optimal state. (In real quantum hardware, a shallow circuit may be less susceptible to noise.) With COBYLA and p = 4 and 5, we safely obtain the desired maxcut, while we don't with p=1 and 2, as shown in Fig. 3.

**Simulation with noise model:** Results when introducing the live noise model and Fake Provider are shown in Fig. 4. Optimization and execution on the Aer simulator yielded the expected dominant bitstrings 00111 and 11000 (Fig. 4(a)). However, the Aer simulator with the live noise model introduced additional noise (Fig. 4(b)). Notably, the Fake Provider yielded results with minimal noise and clearly distinguishable optimal bitstrings (Fig. 4(c)). Results from the IBM hardware will be explained in detail in the next section, but Fig. 4(d) is included for comparison. The results exhibited bitstrings similar to the live noise model in Fig. 4(b) but with increased noisy values. In the next section, we will show that we could reduce the amount of noise and get a more prominent selection of the two answer bitstrings by using error mitigation methods.

Comparing Fig. 4(a) and 4(c), the Fake Provider offers a better result. It looks counterintuitive, however, by incorporating device calibration data, which introduces biases in gate and measurement errors, these imperfections can sharpen the output distribution toward the optimal solution. Thus, the Fake Provider makes the results appear cleaner than the purely ideal Aer simulation. In contrast, the noiseless Aer simulation preserves the full spread of the variational distribution, which is naturally less concentrated on the optimal solution.

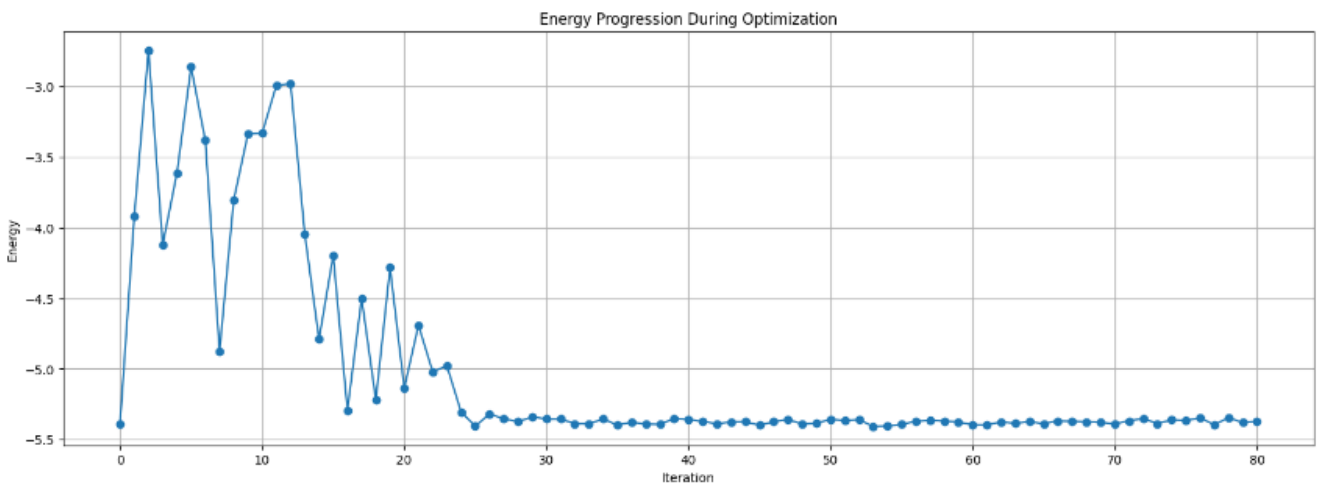

(a) COBYLA (It optimizes the parameters in 81 iterations.)

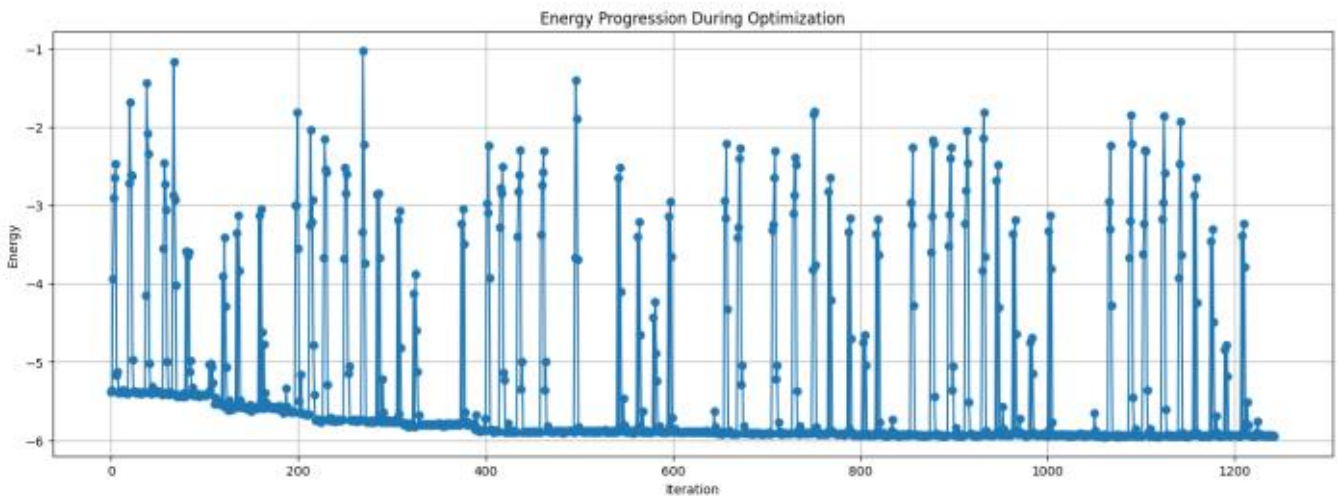

(b) Powell (It optimizes the parameters in 1243 iterations.)

Fig. 2. Energy Progression for QAOA Optimization with COBYLA and Powell and p=5: With COBYLA and Powell, it reaches -5.35 and -6, respectively, while the optimal energy is -6, or equivalently the maximum cut is 6.

### C. *IBM Backend Results*

**COBYLA with circuit depth p:** Similar to the simulator trials, we tested different optimizing methods with a real quantum backend, which led to very interesting results in comparison to the Aer simulator. Fig. 5 displays the bitstrings obtained with COBYLA and p ranging from 2-5. Decreasing the value of p reduced the number of function evaluations

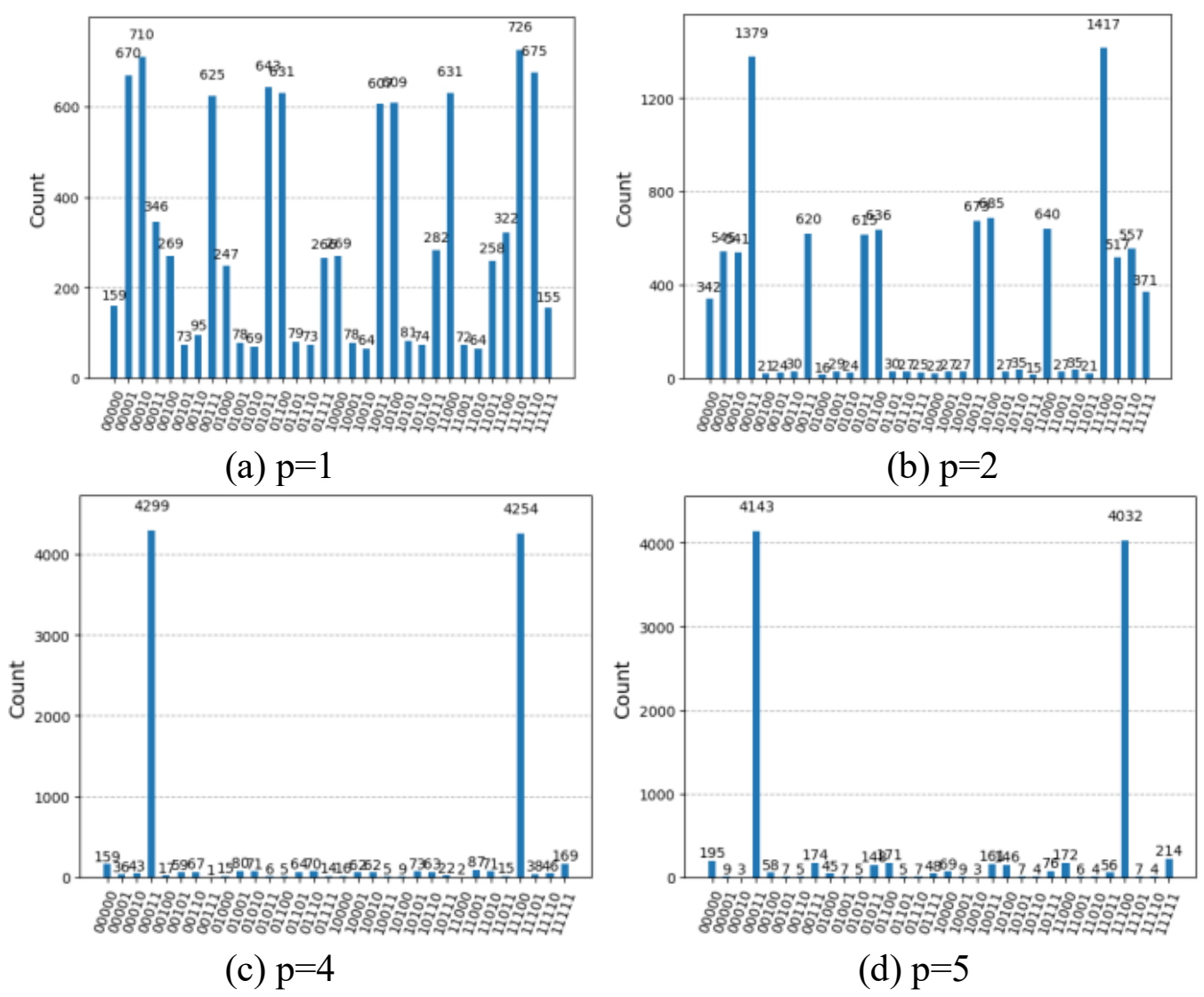

(a) p=1 (b) p=2

(c) p=4 (d) p=5

Fig. 3. Counts (bitstrings) obtained with different p: The number of iterations for the adiabatic process to bring the system to the ground state. As the value of p increases, the probabilities of the two solution cases, 00011 and 11100, become more distinguishable.

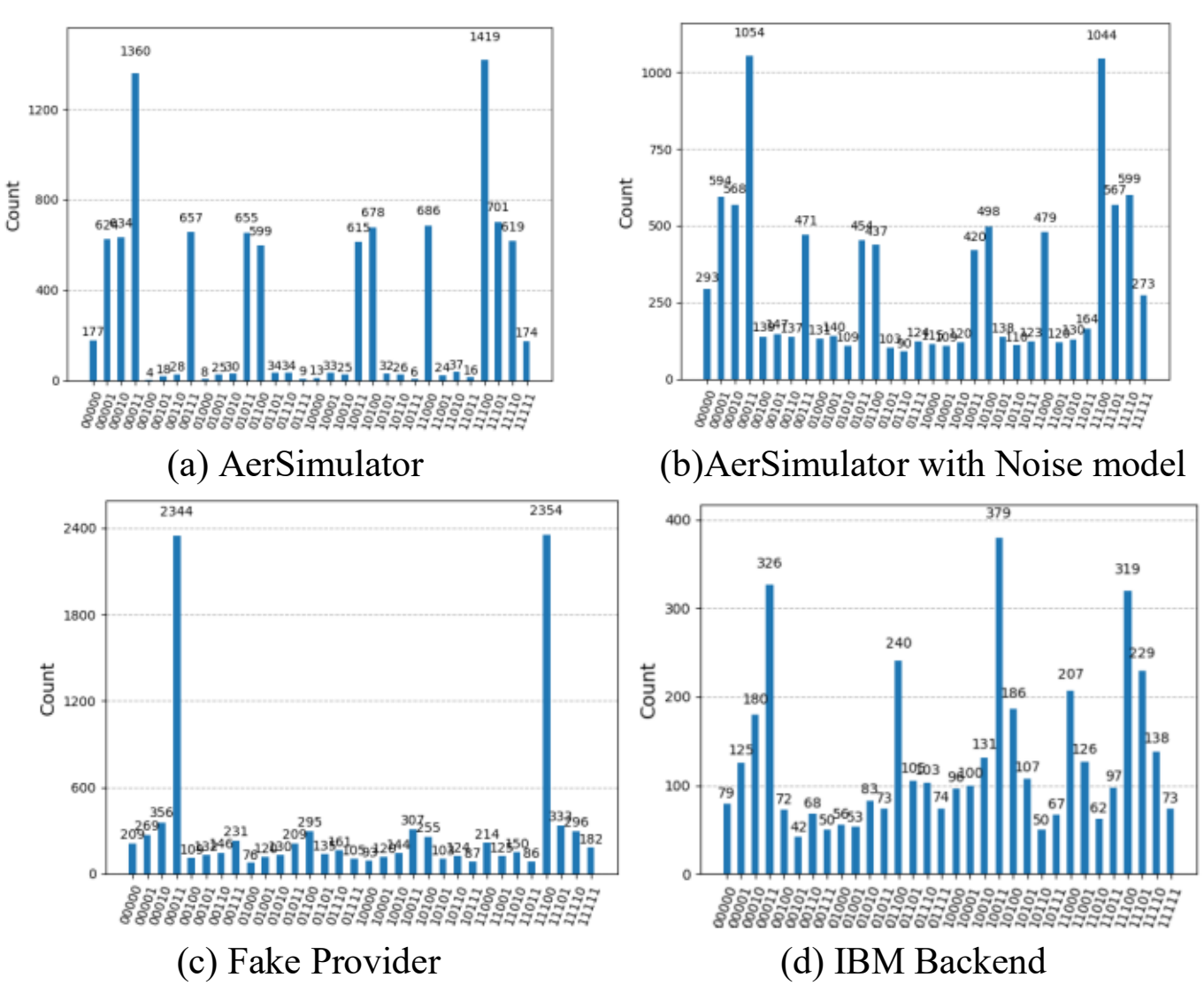

(a) AerSimulator (b)AerSimulator with Noise model

(c) Fake Provider (d) IBM Backend

Fig. 4. Bitstrings obtained from the QAOA circuit. With p = 2 and COBYLA, the circuits were run in a combination of simulator/provider and real quantum hardware. (Fig. 4(a) is the same as Fig. 3(b) but repeated here for comparison. Note also that Fig. 4(c) was based on statistical noise data from *ibm_brisbane* while Fig. 4(d) was from *ibm_cleveland*.)

(iterations) required by the QAOA algorithm to optimize the parameters.

**COBYLA, Powell, and CG**: With COBYLA, when p = 5, it optimized the parameters in 86 iterations. Compared to the simulator results, we observed a noisier outcome, with different bitstrings being explored at higher frequencies, as shown in Fig. 6(a).

As for Powell, it optimized the parameters in 146 iterations with p = 5. Compared to COBYLA, Powell's optimized parameters were more susceptible to noise, as shown in Fig. 6(b). The CG algorithm surprisingly had better results. With p = 5, it took 225 iterations to optimize the parameters. The bitstrings obtained are shown in Fig. 6(c). We can see that the 00011 and 11100 bitstrings were pursued more frequently in the solution space with less noise compared to Powell and COBYLA. The CG method outperformed COBYLA and Powell by achieving better final results, despite requiring more iterations. This can be attributed to CG's ability to more precisely navigate smooth objective landscapes, making it effective at fine-tuning parameters for high-quality solutions [4]. While COBYLA and Powell converged faster, they were more prone to getting trapped in suboptimal regions, especially under the noise and constraints of NISQ devices.

We wanted to observe how the parameters β and γ evolve as the optimization proceeds. Fig. 7 shows that the parameters tend to converge towards certain values as the optimization progresses, with distinct patterns for different iteration ranges, indicating the algorithm's tendency to stabilize over time.

During the convergence of the Quantum Approximate Optimization Algorithm (QAOA), a phenomenon known as parameter plateauing was observed. Specifically, during the parameter optimization process, the parameters reached optimal values and remained unchanged over multiple iterations. Despite this, the optimizer did not stop immediately and continued running for several more iterations before stopping. This behaviour can be explained by the slight improvement in the cost function (Energy Value). Even with fixed parameters, minor improvements in the expectation value can occur due to sampling noise and statistical variation, letting the optimizer proceed until these improvements fall below a certain threshold. Optimizers rely on changes in the cost function rather than parameter changes as their stopping criterion; as long as the cost continues to improve above the threshold, it will keep iterating. QAOA cost landscapes are known to be non-convex and often contain flat plateaus, where gradients are nearly zero but small cost improvements are still possible. Even when the parameters do not change, the quantum circuit continues to execute the same rotations repeatedly, effectively exploring the optimal region until the cost does not improve anymore.

### *D. Results with Error Mitigation Techniques*

We tested the graph shown in Fig. 1(a) using QAOA with error mitigation methods described in Section II.E, namely, dynamic decoupling with the XY4 sequence and Pauli twirling. Additionally, we implemented the transpiler settings of *optimization_level=3* to select the best choice of gates and available qubits. With the error mitigation methods used on IBM hardware, the results, as shown in Fig. 8(b), exhibited more accurate bitstrings compared to those without error mitigation methods in Fig. 8(a). By using error mitigation methods, we could reduce the amount of noise and get a more prominent selection of the two answer bitstrings.

Q-CTRL's Fire Opal quantum solver [17], introduced in Section II, integrates error mitigation methods and additional techniques and was compared. As shown in Fig. 8(c), it exhibited reduced amounts of noise in comparison when executed on an IBM Backend due to its advanced error suppression pipeline.

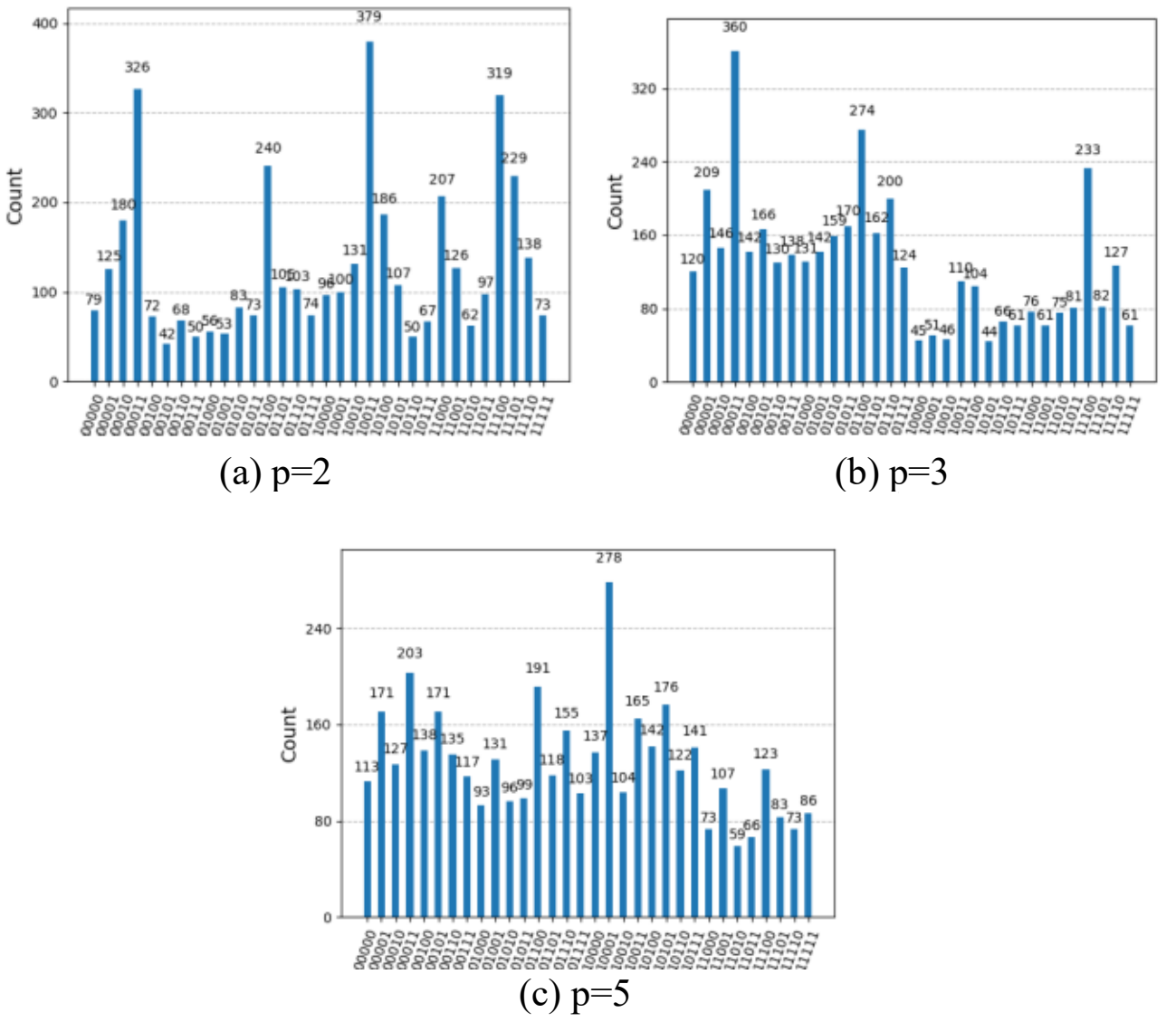

Fig. 5. Counts (bitstrings) obtained with different p on **IBM Backend** with COBYLA: It optimizes the parameters in 55, 67, and 86 iterations, respectively. It is observed that a higher p does not necessarily provides a better result. (Note that Fig. 5(a) is the same as Fig. 4(d) but repeated here for comparison.)

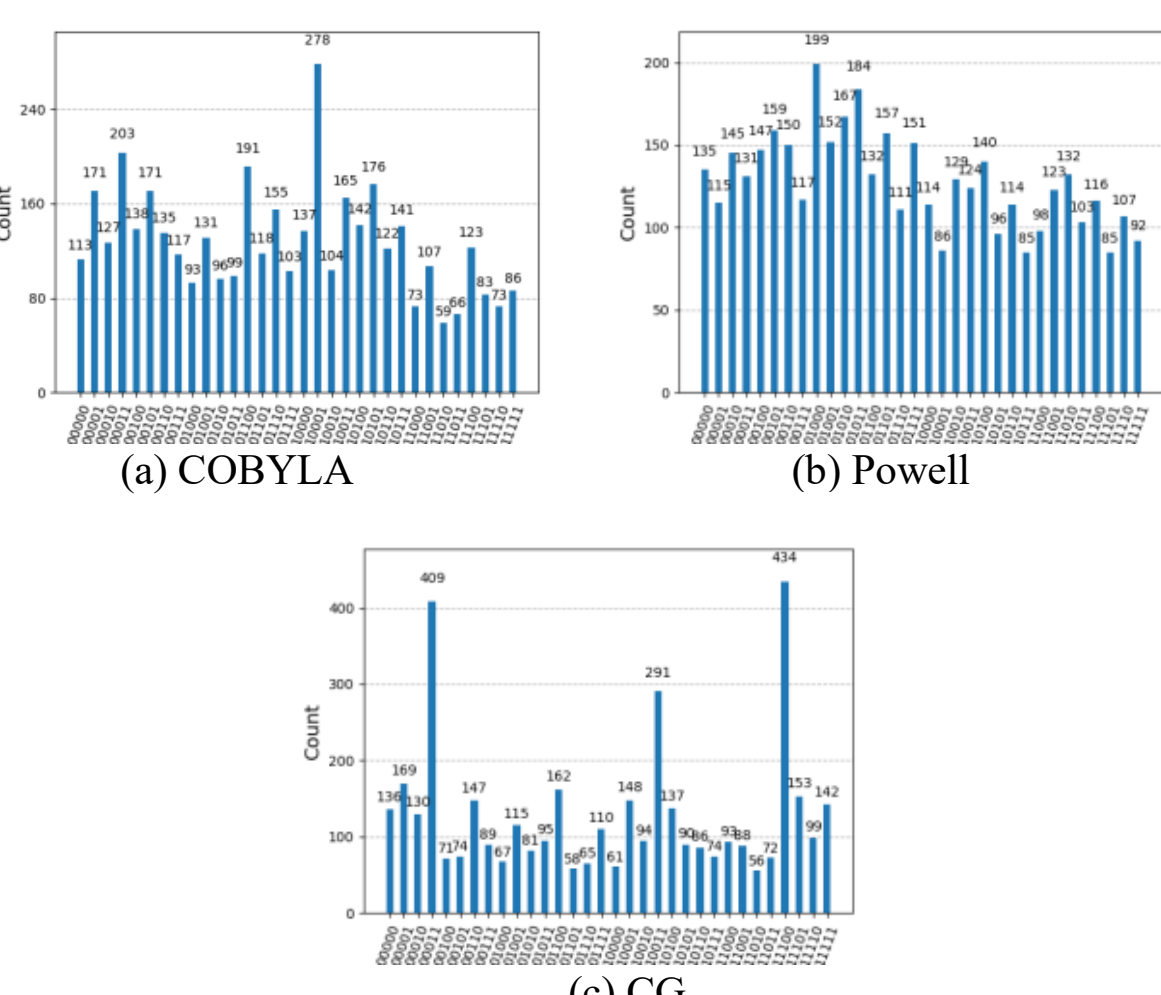

Fig. 6. Bitstrings obtained from QAOA on **IBM Backend** (p=5; The results are not as good as in simulations, shown in Fig. 5. The CG algorithm, shown in (c), surprisingly shows the best performance, from our experiments.

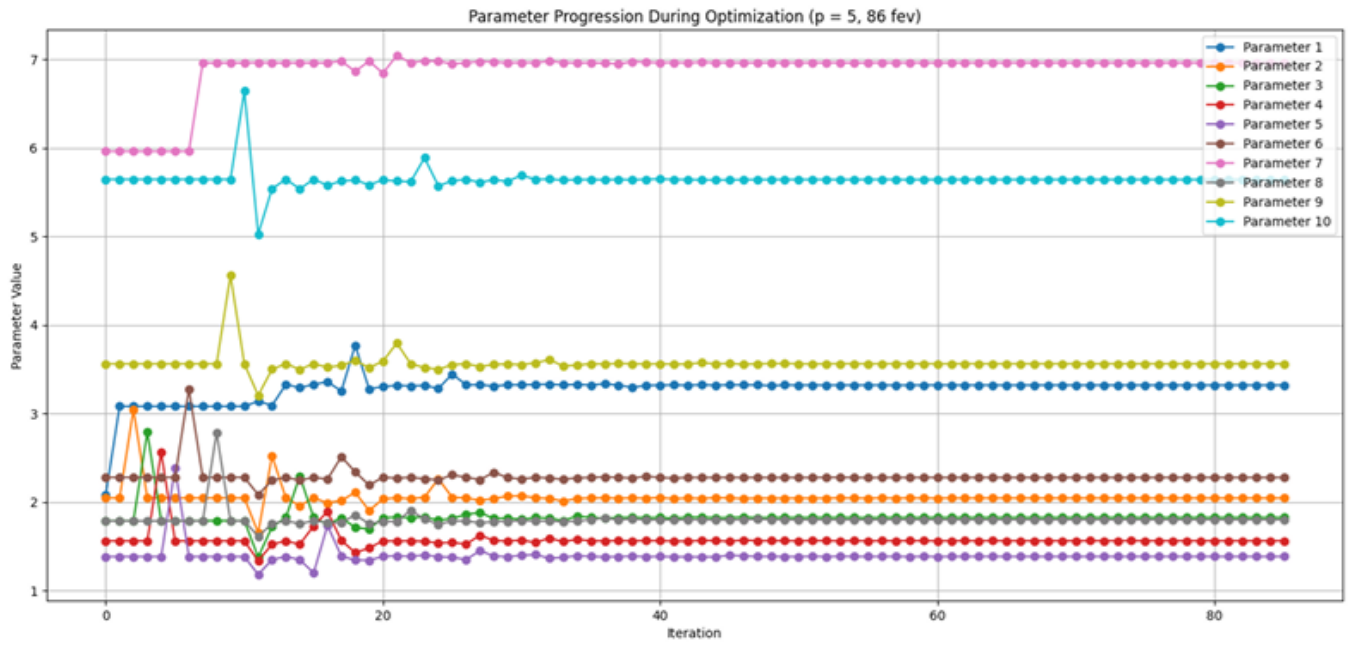

(a) COBYLA (It optimizes the parameters in 86 iterations.)

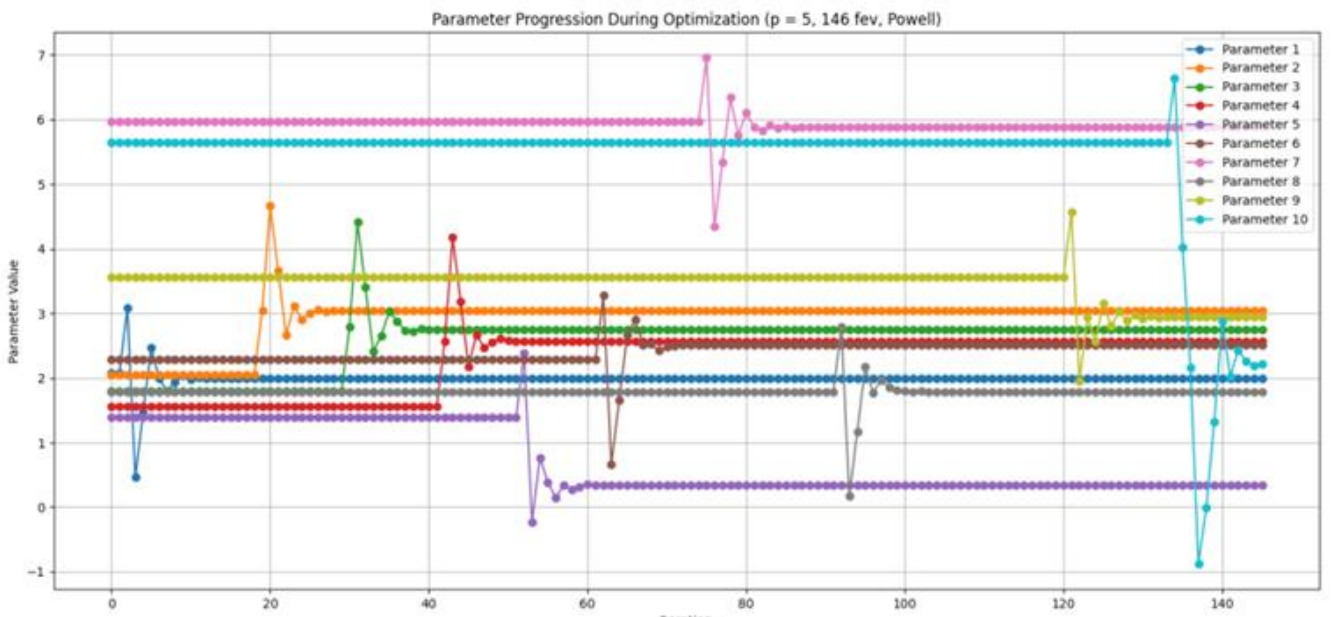

(b) Powell (It optimizes the parameters in 146 iterations.)

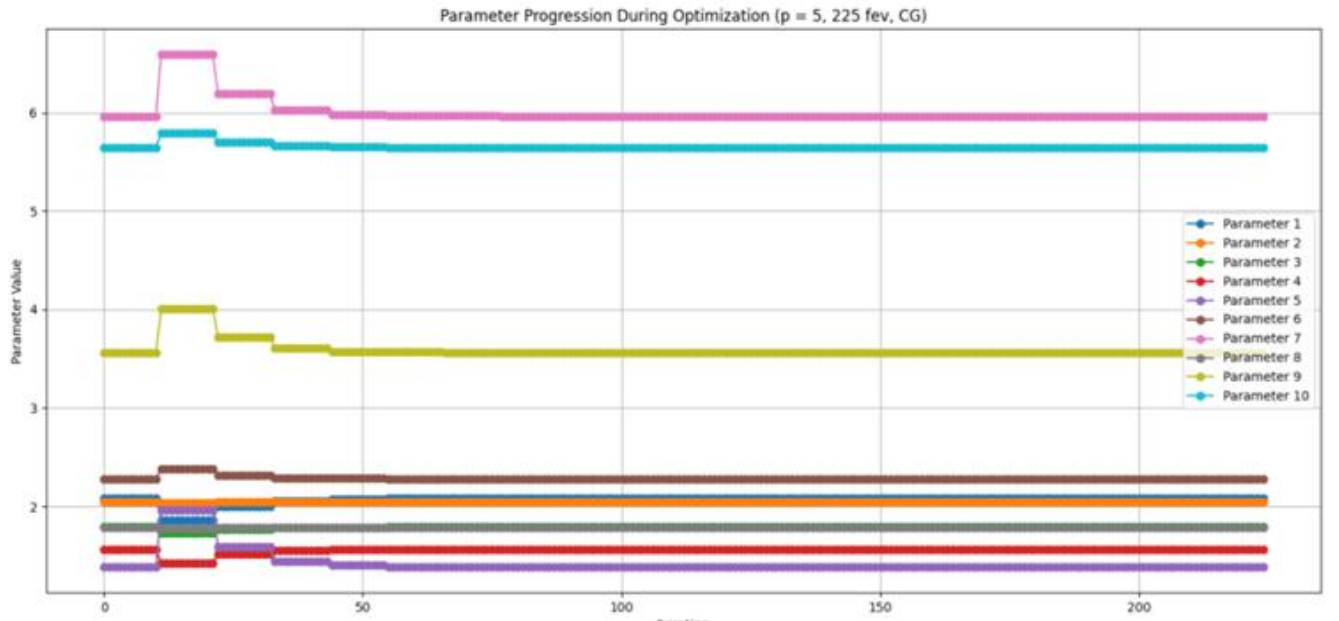

(c) CG (It optimizes the parameters in 225 iterations.)

Fig. 7. Parameter Progression for QAOA Optimization on **IBM Backend** with p=5: Fig. (a) shows how the parameters initially fluctuate but stabilize after 20 iterations with COBYLA; Fig. (b), Powell's parameter progression appears erratic and fluctuates significantly throughout the optimization process; Fig. (c) shows CG's parameter progression stabilizes relatively early on.

In Fig. 8(c), Fire Opal exhibits the highest probability for 11000, which is equivalent to 00011, as it does not invert bitstrings. Moreover, Fire Opal gives us only one of the two solutions because it suppresses its complementary partner due to its "symmetry-breaking" effects [16]. In other words, Fire Opal's optimizer trajectory, device/readout biases, or sampling noise break the symmetry, causing asymmetric sampling of degenerate solutions.

### E. Analysis and Limitations

For the simulator, the optimization method or the value of p does not matter as much, as we almost always get near-perfect results. By decreasing p, there is a chance that the simulator will explore the wrong solution space, but the probability of that is minimal.

When looking at the results of the real quantum hardware, the COBYLA and CG optimization methods gave the best results. CG, even though it took more iterations in comparison to Powell, gave better results overall. As the value of p decreased, the optimization process became easier and took less time and iterations since the layers of the circuit decreased as well. It was not as computationally intensive for the quantum computer to carry out.

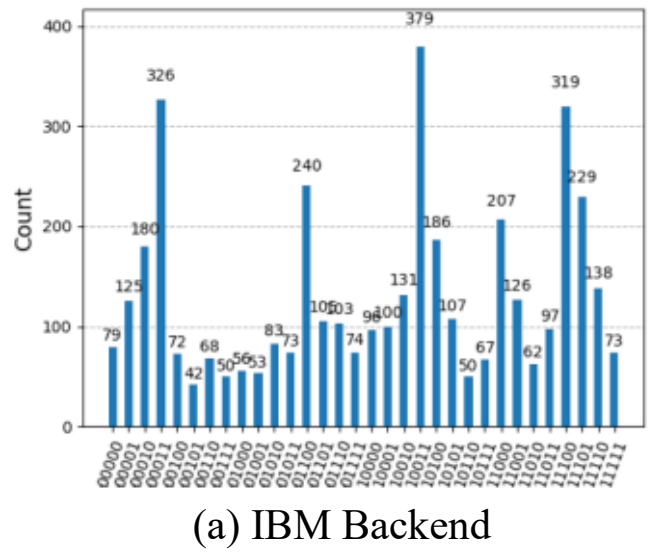

(a) IBM Backend

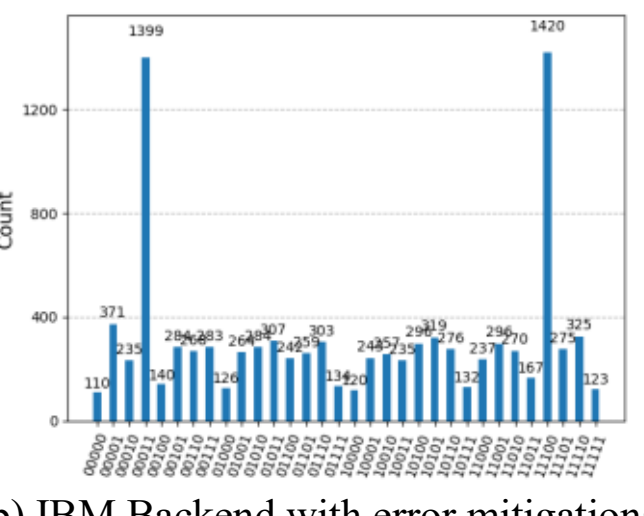

(b) IBM Backend with error mitigations

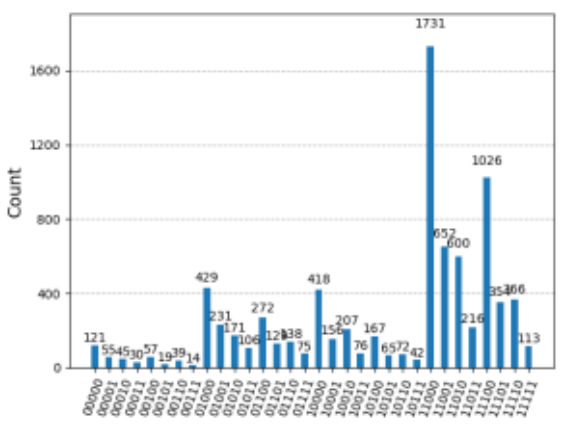

(c) IBM Backend with Fire Opal

Fig. 8. Bitstrings obtained from the QAOA circuit with p = 2 and COBYLA: Fig. 8(a) is the same as Fig. 4(d) but repeated here for comparison. We used *ibm_cleveland* for (a) and (b) and *ibm_sherbrooke* for (c).

When running on an IBM backend, we integrated error mitigation strategies, including Pauli twirling, dynamic decoupling (XY4 sequence), and transpiler optimizations (optimization_level=3). This gave us results comparable to Q-CTRL's Fire Opal QAOA solver that improved performance at the pulse and gate level.

**Live Noise Model and Fake Providers:** Fake Providers (e.g., *FakeSherbrooke* & *FakeBrisbane*) are simulated representations of real IBM hardware generated from historical calibration data. They serve as static snapshots of device properties, including coupling maps, gate error rates, and relaxation/decoherence times ($T_1/T_2$), making them valuable for transpilation testing, hardware-aware compilation, and noise evaluation without live QPU access. In contrast, Aer simulators allow noise models to be constructed from current backend data (from_backend) or defined manually, providing more flexible and scalable noise simulations. Both of these complement real hardware execution, where circuits are subject to live quantum noise, device connectivity constraints, and hardware latency.

In this work, we employed realistic noise models by running QAOA circuits on Aer simulators configured from actual IBM backends, as well as on IBM's Fake Providers. In comparing execution environments, we observed notable differences between Aer simulators (with and without live noise models) and Fake Providers. Results from Fake Providers were strikingly clean, with little noise and sharply defined results. However, note that the purpose of both the noise model and the Fake Provider is to provide a realistic runtime environment rather than offering a cleaner solution.

In comparison, AerSimulator using live noise models produced substantially noisier results. While this reflects their design to inject stochastic noise consistent with calibration data, it often overestimates the impact of noise compared to actual hardware runs. The difference may come from simplifications in how Aer models correlated errors, crosstalk, and temporal drift. AerSimulator with live noise models may be more useful for stress-testing algorithm robustness.

## V. Conclusion

In this paper, we explored the effectiveness of the Quantum Approximate Optimization Algorithm for solving the Maxcut problem of a graph with 5 nodes. COBYLA and CG were found to be effective methods for the QAOA circuit under the experimental environment we tested. These methods helped achieve better performance regarding convergence to a good solution. Decreasing p, the depth of the circuit, decreases the computation time. However, this reduction came at the potential cost of solution quality, as p controls the number of alternating operators applied, which impacts the algorithm's ability to explore the solution space.

We observed key differences between results on the simulator and real quantum hardware, most notably, the Powell method's unique energy progression trend in Fig. 3(b), which deviated significantly from quantum hardware results in Fig. 6(b). Similarly, increasing the circuit depth p did not consistently lead to better performance as shown in Fig. 5. Fig. 7 also revealed distinct fluctuations in parameter progression across different optimization methods, which merit deeper investigation.

This study explored QAOA across simulators, Fake Providers, and IBM hardware, incorporating error mitigation techniques such as Pauli twirling, dynamic decoupling, transpiler optimizations, and Q-CTRL's Fire Opal platform. Our experiments highlighted the practicality of hybrid workflows, where optimization is performed on simulators or Fake Providers and final circuits are executed on hardware, yielding consistent and reliable results.

While Fire Opal provides a commercial and proprietary solution that abstracts away much of its internal methodology, we demonstrated that transparent, reproducible techniques, clearly stating the methods and parameters used, can yield comparable results. In this way, our work provides both validation of existing platforms and a path toward more open, customizable approaches to error mitigation and circuit optimization in QAOA.

Moving forward, we aim to explore the causes of QAOA discrepancies in greater detail and evaluate additional optimization strategies for QAOA. We also plan to experiment with alternative ansatzes to determine their impact on performance. Along with a more thorough analysis of IBM hardware's qubit connectivity and constraints, this will help guide better qubit selection, potentially improving QAOA execution on noisy quantum devices. Finally, we aim to use the QOBLIB – Quantum Optimization Benchmarking Library, which provides challenging problem classes with practical relevance and varying complexity, to systematically evaluate QAOA performance on instances that are difficult for classical methods and span system sizes up to 10,000 variables [14].

## Acknowledgment

We want to express our sincere gratitude to the Cleveland Clinic for their generous support in allowing us to use the *ibm_cleveland* quantum system during the summer of 2024.

## References

[1] E. G. Rieffel and W. Polak, "An Introduction to Quantum Computing for Non-Physicists," *arXiv*, quant-ph/9809016, 2000. [Online]. Available: https://arxiv.org/abs/quant-ph/9809016. [Accessed: Apr. 2, 2025].

[2] C. Outeiral, M. Strahm, J. Shi, G. M. Morris, S. C. Benjamin, and C. M. Deane, "The prospects of quantum computing in computational molecular biology," *WIREs Computational Molecular Science*, vol. 11, no. 1, May 2020, doi: https://doi.org/10.1002/wcms.1481.

[3] L. Zhou, S.-T. Wang, S. Choi, H. Pichler, and M. D. Lukin, "Quantum Approximate Optimization Algorithm: Performance, Mechanism, and Implementation on Near-Term Devices," Physical Review X, vol. 10, no. 2, Jun. 2020, doi: https://doi.org/10.1103/physrevx.10.021067.

[4] K. Blekos, D. Brand, A. Ceschini, C.-H. Chou, R.-H. Li, K. Pandya, and A. Summer, "A review on Quantum Approximate Optimization Algorithm and its variants," *Physics Reports*, vol. 1068, pp. 1-66, Jun. 2024. [Online]. Available: http://dx.doi.org/10.1016/j.physrep.2024.03.002. [Accessed: Apr. 2, 2025].

[5] "Quantum technology | IBM Quantum Learning," *Ibm.com*, 2020. https://learning.quantum.ibm.com/course/quantum-business-foundations/quantum-technology (accessed Mar. 21, 2025)

[6] R.Shaydulin, "QAOA Tutorial Hands-on Notebook", 2023, Github repository. [Online]. Available: https://github.com/rsln-s/QAOA_tutorial/blob/main/Hands-on.ipynb.

[7] IBM Quantum Learning, Variational Quantum Algorithms, https://learning.quantum.ibm.com/course/variational-algorithm-design/variational-algorithms

[8] Cerezo, M., Arrasmith, A., Babbush, R., Benjamin, S.C., Endo, S., Fujii, K., McClean, J.R., Mitarai, K., Yuan, X., Cincio, L. and Coles, P.J., 2021. Variational quantum algorithms. *Nature Reviews Physics*, *3*(9), pp.625-644.

[9] A. J. et al., "Quantum Algorithm Implementations for Beginners," *ACM Transactions on Quantum Computing*, vol. 3, no. 4, pp. 1-92, Jul. 2022. [Online]. Available: http://dx.doi.org/10.1145/3517340. [Accessed: Apr. 2, 2025]

[10] "IBM Quantum," *IBM Quantum*, 2025. https://quantum-computing.ibm.com/services/resources?tab=systems. (accessed Feb. 10, 2025).

[11] IBM, "IBM Quantum," https://www.ibm.com/quantum/qiskit.

[12] IBM Quantum Documentation, "Transpiler | IBM Quantum Documentation", 2024. [Online]. Available: https://docs.quantum.ibm.com/api/qiskit/transpiler#layout-stage.

[13] G. W. Dueck, A. Pathak, M. M. Rahman, A. Shukla, and A. Banerjee, "Optimization of Circuits for IBM's Five-Qubit Quantum Computers," in *2018 21st Euromicro Conference on Digital System Design (DSD)*, Aug. 2018, pp. 680-684. [Online]. Available: http://dx.doi.org/10.1109/DSD.2018.00005. [Accessed: Apr. 2, 2025].

[14] "QOpt / QOBLIB - Quantum Optimization Benchmarking Library · GitLab," *GitLab*, 2025. https://git.zib.de/qopt/qoblib-quantum-optimization-benchmarking-library (accessed Apr. 12, 2025).

[15] QPU Information, IBM, https://docs.quantum.ibm.com/guides/qpu-information (accessed Apr. 14, 2025).

[16] N. Sachdeva *et al.*, "Quantum optimization using a 127-qubit gate-model IBM quantum computer can outperform quantum annealers for nontrivial binary optimization problems," *arXiv (Cornell University)*, Jun. 2024, doi: https://doi.org/10.48550/arxiv.2406.01743.

[17] "Automated performance optimization for quantum algorithms | Q-CTRL," *Q-ctrl.com*, 2025. https://q-ctrl.com/fire-opal (accessed Sep. 28, 2025).